# Dependency of heat transfer rate on the Brinkman number in microchannels

Hee Sung Park

Stokes Institute, University of Limerick,

Limerick, Ireland

**Abstract**

Heat generation from electronics increases with the advent of high-density integrated circuit technology. To come up with the heat generation, microscale cooling has been thought as a promising technology. Prediction of heat transfer rate is crucial in design of microscale cooling device but is not clearly understood yet. This work proposes a new correlation between heat transfer rate and Brinkman number which is nondimensional number of viscosity, flow velocity and temperature. Our experimental results showed a good empirical equation that $Nu/(Re^{0.62} Pr^{0.33})$ is inversely proportional to the Brinkman number in laminar flow regime. It is expected that the equation proposed by this work can be useful to design microchannel cooling device.

**Introduction**

Microscale fluidic technologies have revolutionized many aspects of applied sciences and engineering, such as microflow sensors, pumps, valves, thin film coating, heat exchangers, combustors, fuel processors, and biomedical and biochemical analysis instruments [1][2]. The use of microchannels has become common in these applications. Many scientific and engineering programmes have been conducted to develop more effective and sophisticated devices. From a practical point of view, prediction of heat transfer rate and friction is significant to design microfluidic device. For example, Park et al. [3] showed an methodology of determining the optimum microchannel dimensions for electronic cooling device. Hetsroni et al. [4] conducted experimentation on microchannel heat transfer rate and found that flow instabilities caused pressure fluctuation which decreased the heat transfer rate. Harm et al. [5] analyzed their experimental data assuming developing flow conditions, concluding that convective heat transfer theory was applicable. The numerical calculations reported by Lee et al. [6] were compared with their experimental data, and it was found that a conventional analysis approach could be employed in predicting heat transfer behavior in microchannels. Tso and Mahulikar [7] investigated the effect of viscous dissipation in microchannel heat transfer and suggested the dependence of $Nu/(Re^{0.62} Pr^{0.33})$ on Brinkman number which originated from Peng et al. [8]. However, large scatter of data from published papers is evident in measuring data for friction factor and heat transfer, and explanations are contradictory [9][10]. Meanwhile, recent papers have indicated that conventional theory is applicable by careful evaluation of geometry [11]. Specifically, many previous experimental studies were conducted by using multiple numbers of channels in test sections without showing the manifold condition. Non-uniform distribution of flow in the multiple channels can cause significant distortion on the analyses of experimental data for friction factor and heat transfer.

In this paper, experimental and theoretical analyses were carried out on a range of microchannel arrays with uniform flow distribution in order to determine accurate correlations for heat transfer rate. A manifold geometry was created to ensure uniform inlet flow for each microchannel, and the friction factor and heat transfer rate were carefully evaluated and then compared with theory. The friction factor obtained by the experiments showed an excellent agreement with conventional hydraulic theory, however, heat transfer rates showed large scattering compared with conventional theory. We analyzed the Nusselt number by the correlation of $Nu/(Re^{0.62} Pr^{0.33})$ and proposed a correlation between heat transfer rate and Brinkman number.

**Theoretical**

The schematic diagram of a representative microchannel heat exchanger is presented in Fig. 1. Heat flux ($q$) is applied from the vertical direction through the upper plate of thickness ($t$). The channel dimensions are expressed with channel width ($D_c$), channel depth ($D_d$), wall thickness ($D_w$) between the channels, and overall width ($W$) and length ($L$).

As the present work concerns the single-phase laminar flow in microchannels, the well-known Fanning friction factor ($C_f$), defined in equation (1) accounts for the microchannel flow friction caused by wall shear stress:

$$C_f = \frac{(\Delta P / L) D_h}{2\rho V^2} \quad (1)$$

$$D_h = \frac{2 D_d D_c}{D_d + D_c} \quad (2)$$

where $\Delta P$ is pressure drop across the channel, $D_h$ is the hydraulic diameter defined in equation (2), and $\rho$, $V$ represent density and mean velocity of the liquid flow respectively. For fully-developed laminar flow, the Poiseuille number is a function of channel aspect ratio, $\alpha = D_w/D_d$ [12][13]. Equation (3) expresses the friction factor in terms of the geometric parameter, $G$, defined by Bejan [14] [15].

$$C_f \operatorname{Re} = 4.7 + 19.64 G \quad (3)$$

$$G = \frac{\alpha^2 + 1}{(\alpha + 1)^2} \quad (4)$$

For the fully-developed flow regions, the pressure drops for the given flow rates across the channel can be evaluated using the calculated friction factor.

The heat transfer coefficient, $h$, which is significant for microscale heat exchanger design, can be obtained from the Nusselt number using the following equation (5).

$$h = \frac{k Nu}{D_h} \quad (5)$$

Assuming a fully-developed temperature profile with constant heat flux, the Nusselt number is expressed in equation (6) as a function of geometric parameter, $G$ [15]. The correlation shows good agreement with the analytical results of Kays and Crawford [13].

$$Nu = -1.047 + 9.326 G \quad (6)$$

When the fluid temperature upstream of some point is assumed to be uniform and equal to the surface temperature, there is no heat transfer in this region [16]. Following this point, the heat transfer is concerned with the development of the temperature profile. In the developing temperature profile, a heat transfer solution can be obtained by defining the thermal entrance length [13]:

$$x^* = \frac{x}{\operatorname{Re} D_h \operatorname{Pr}} \quad (7)$$

where $x^*$ and Pr represent thermal entrance length and Prandtl number, respectively. The average Nusselt number obtained by fitting the data from Shah and London [17] is shown in [16]:

$$Nu = \left[ \{2.22 (x^*)^{-0.33}\}^3 + \{-0.02 + 8.31 G\}^3 \right]^{1/3} \quad (8)$$

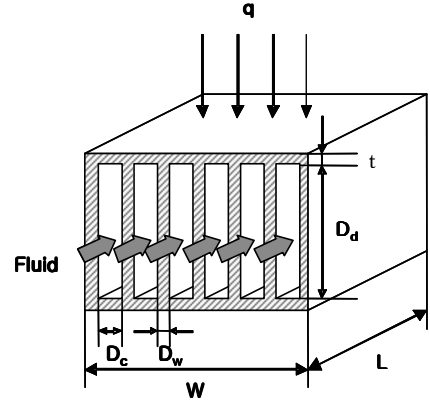

Fig. 1. Schematic of microchannel heat exchanger

It is evident from the literature that although conventional theory predicts friction factor accurately, significant discrepancies exist for heat transfer. In this paper, a novel heat transfer correlation is proposed, based on experimental data measured for microchannel arrays with uniform flow distribution.

**Experiments**

A range of rectangular shaped microchannels were fabricated by chemical wet etching <110> silicon wafer of 500 μm thickness. The silicon wafer was then anodically bonded with 500 μm Pyrex glass plate in order to minimize heat loss. A Pt/Ti thin film heater layer of 0.4 μm thickness was deposited on the backside of the silicon wafer. The total electrical resistance of the heater was 190 Ω. Finally, a black lacquer was coated over the thin film layer to measure the temperature profile using infra-red thermography. Finally, each of the microchannel arrays was cut to 10 mm X 10 mm to ensure the same effective heating area. At the completion of testing, the microchannel arrays were cross-sectioned in order to examine the dimension and shape using optical microscopy and we found out there were no significant undercuts or unexpected deformations. The analysis based on the rectangular shaped microchannel is thought to be valid. The measured microchannel dimensions are presented in Table 1 where the microchannels were classified into two groups: the channel widths of the first group were varied from 65 μm to 315 μm, while the second group had same channel width of 100 μm but with varying numbers of channels. Each inlet and outlet of microchannels was assembled with manifold parts made of Teflon, designed to sustain uniform flow distribution for each microchannel. A schematic diagram of the experimental setup is shown in Fig. 2. It consisted of a liquid filter, liquid mass flow controller, test unit, power supply, infra-red camera, data acquisition system, and a liquid circulator included a liquid pump and liquid temperature controller unit. Deionized water was used for all the experiments and a liquid filter was placed front of the liquid circulator to prevent contaminants from flowing into the test section. The flow rate was controlled by the liquid mass flow controller. The measurement data from the pressure sensor and the infra-red camera were collected by data acquisition system. During experimentation, each microchannel heat exchanger assembled with manifold parts (Fig. 3) was connected using inlet and outlet tubes. After setting the desired flow rate and temperature of water, electrical current was supplied to the thin film heater. When all the signals displayed on the data acquisition system reached steady-state, the experimental data were recorded. The actual supplied power was calculated from the temperature difference measured by thermocouples between inlet and outlet fluids and compared with the power from measuring the electric conditions.

Table 1. Configurations of tested microchannels

| (μm) | 1 | 2 | 3 | 4 | 5 | 6 | A | B | C | D |
|---|---|---|---|---|---|---|---|---|---|---|
| $D_c$ | 65 | 114 | 166 | 216 | 265 | 315 | 100 | 100 | 105 | 104 |
| $D_w$ | 35 | 88 | 133 | 177 | 248 | 292 | 20 | 40 | 75 | 137 |
| $D_h$ | 106 | 165 | 214 | 251 | 281 | 307 | 150 | 150 | 156 | 154 |
| $N_c$ | 100 | 49 | 33 | 25 | 19 | 16 | 83 | 71 | 55 | 41 |
| Re | 69~218 | 124~392 | 194~519 | 230~616 | 230~742 | 247~807 | 77~241 | 89~281 | 113~358 | 151~480 |

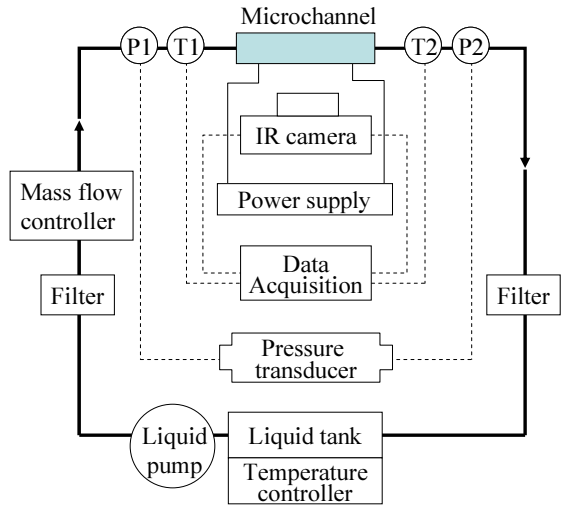

Fig. 2. Schematic diagram of test apparatus

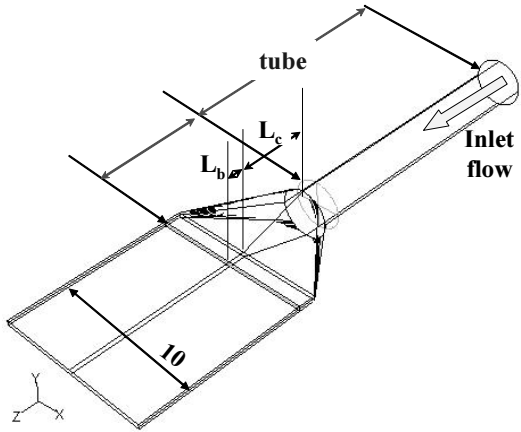

Fig. 3. Manifold structure for uniform flow distribution in the microchannel array

An uncertainty analysis was carried out on relevant parameters using the procedure described by Holman and Gajda [18]. In this paper, the maximum values of uncertainty were dominated by the flow controller, manometer, and temperature measurement limits. The uncertainties in Reynolds number, friction factor, and Nusselt number were 2.0 %, 9.2 %, and 8.7 %, respectively.

## Results and Discussions

The inlet and outlet manifolds are significant in measuring pressure drop and heat transfer in multiple microchannels flows [5]. The pressure drop in the manifolds as a fractions of the measured pressure drops in the microchannel ($\Delta P_{mani}/\Delta P_{exp}$) was typically below 17 % as shown in this work and the fraction was less than 10 % in 33 data points out of a total 40 data points. In the experiments, the measured pressure data was corrected by subtracting the pressure drop of the manifolds. The friction factors were evaluated from the experimental pressure drop data by using equation (1) that are plotted in Fig. 4. Regarding the curve, experimental data were consistent with the values predicted from the theory for fully-developed flow.

The average Nusselt number was determined by using Newton's law of cooling from the measurement data:

$$h = q / A(T_w - T_m) \qquad (9)$$
$$A = nL(D_c + 2D_d) \qquad (10)$$

A represents the fin area of cooling surface inside channels. $T_w$ is silicon wall temperature wetted by water and this is estimated by the integrating the solution of the fin equation [10]. $T_m$ (mean water temperature) is evaluated from averaging inlet ($T_i$) and outlet temperatures ($T_o$).

$$T_m = \frac{1}{2}(T_o - T_i) \quad (11)$$

The average Nusselt numbers with respect to Reynolds number from the experiments and theory are compared in Fig. 5. It shows that the experimental data are scattered between the curves plotted by the theories with the assumptions of fully-developed and developed temperature profile. As the equations from (9) to (11) are empirical relationships obtained from macroscale flow conditions, the inconsistency is expected to be due to microscale effects. The curves of Fig. 5 show different relations for each flow rate. In the heat transfer experiments, constant heat flux was applied, and then the mean water temperature on the range of the flow rates varied from 28 °C to 40 °C. In this range of temperature variation, the viscosity varied 18 % whereas other thermophysical properties (conductivity, heat capacity, and density) of water vary by less than 2 %. It can be deduced that flow velocity and viscosity variations cause microscale effect on the Nusselt number. In this respect, the Brinkman number is thought to be a characteristic nondimensional parameter for accounting the microscale effects in configurations specified in terms of heat-flux [13]. In this regard, Tso and Mahulikar [7] proposed the relationship of Brinkman number with $Nu / (Re^{0.62} Pr^{0.33})$. The combined parameter of $Nu / (Re^{0.62} Pr^{0.33})$ originated from Peng et al. [8] by analyzing their experimental data and by modifying the Sieder-Tate equation [19]. They proposed a new correlation of $Nu / (Re^{0.62} Pr^{0.33})$ which features an empirical constant depending on the geometric configuration. Tso and Mahulikar [7] correlated the combined parameter with the Brinkman number in order to explain unusual behavior in the Nusselt number. Consequently, our experimental data were plotted again in Fig. 6 using the new correlation and a relationship with Brinkman number was found within Prandtl number and flow velocity ranges of 4.44 ~ 5.69 and 0.34 m/s ~ 2.61 m/s, respectively. The correlation was expressed in equation (12) by curve fitting from our experimental data.

$$Nu / (Re^{0.62} Pr^{0.33}) = 0.015 \, Br^{-0.22} \quad (12)$$

It showed good agreement for samples A to D – those with the same hydraulic diameter of 150 μm, with different number of channels, i.e. different flow velocities in the channels. Our analysis on the

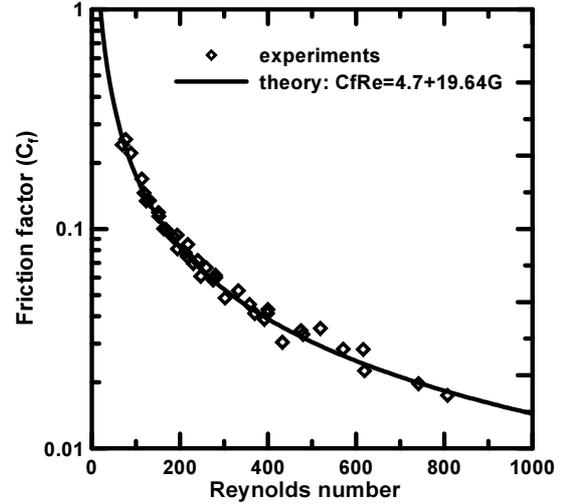
Fig. 4. A good comparison between theory and experiment

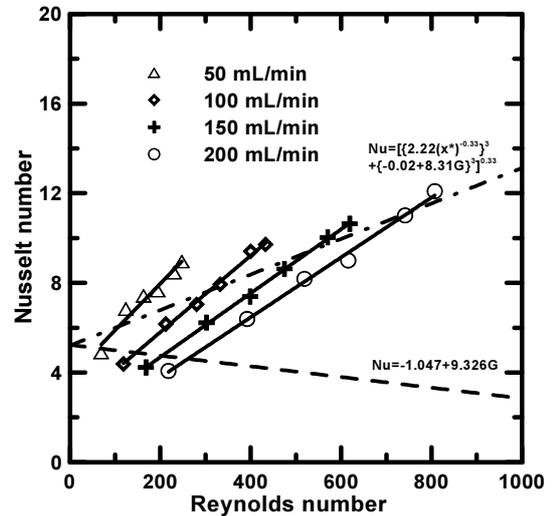
Fig. 5. Nusselt number as a function of Reynolds number

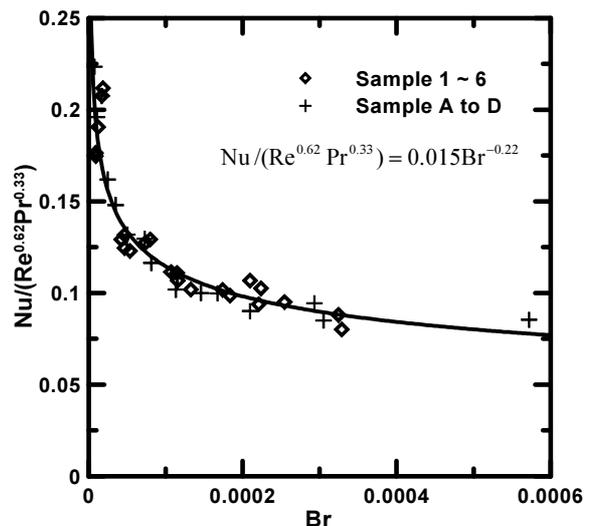
Fig. 6. $Nu/(Re^{0.62}Pr^{0.33})$ as a function of Br proposed in this paper

experimental results for the heat transfer suggests that the Nusselt number can be predicted with accuracy using Brinkman number. Even though the correlation is related to Brinkman number – a parameter for viscous dissipation – the contribution of viscous dissipation to the temperature difference between mean water and channel wall is negligible because of the low-Prandtl number with low speed flow in our experiments. Hetsroni et al. [10] also concluded that the effect of viscous dissipation on heat transfer in microchannel is negligible under typical flow conditions. From the definition of Brinkman number, it implies the ratio of the momentum transfer ($\mu V^2$) to the conduction heat transfer ($k\Delta T$) across the flow. It is thought that the correlation of $Nu/(Re^{0.62} Pr^{0.33})$ and Brinkman number indicates the presence of some microscale effects on the heat transfer induced by variations of viscosity and flow velocity in the microchannel.

**Conclusions**

The heat transfer rates in a range of microchannel arrays were experimentally investigated. Our experimental results deviated from the Nusselt number calculated from conventional heat transfer theory but were analyzed by using the correlation of $Nu / (Re^{0.62} Pr^{0.33})$ and Brinkman number. We proposed a new correlation of heat transfer in microchannels on this basis.